\documentclass[fleqn,useAMS,usenatbib,usegraphicx]{mn2e}
\usepackage{amsmath}
\usepackage{amssymb}
\usepackage{times}
\usepackage{booktabs}

\newcommand{\Msun}{\hbox{$\,\text{M}_{\sun}$}}
\newcommand{\Rsun}{\hbox{$\,\text{R}_{\sun}$}}
\newcommand{\Lsun}{\hbox{$\,L_{\sun}$}}

\newcommand{\Mbh}{M_\text{BH}}

\newcommand{\stars}{\textsc{stars}}
\newcommand{\uvp}{$U$--$V$ plane}
\newcommand{\pdif}[2]{{\frac{\partial #1}{\partial #2}}}
\newcommand{\tdif}[2]{{\frac{\mathrm{d} #1}{\mathrm{d} #2}}}

\newcommand{\shorteq}[1]{
\begin{equation}
#1
\end{equation}}
\newcommand{\mchead}[1]{\multicolumn{2}{c}{#1}}

\newcommand{\aap}{A\&A}

\newcommand{\apj}{ApJ}

\newcommand{\apss}{Ap\&SS}

\newcommand{\mnras}{MNRAS}

\begin{document}

\title[Quasi-stars, giants and the SC limit]
{Quasi-stars, giants and the Sch\"onberg--Chandrasekhar limit}
\author[W. H. Ball et al.]  {Warrick H. Ball\thanks{E-mail:
    wball@ast.cam.ac.uk},
  Christopher A. Tout, and Anna N. \.Zytkow\\
  Institute of Astronomy, The Observatories, Madingley Road,
  Cambridge CB3 0HA}
\maketitle

\begin{abstract}
  The Sch\"onberg--Chandrasekhar (SC) limit is a well-established
  result in the understanding of stellar evolution. It provides an
  estimate of the point at which an evolved isothermal core embedded
  in an extended envelope begins to contract. We investigate contours
  of constant fractional mass in terms of homology invariant variables
  $U$ and $V$ and find that the SC limit exists because the isothermal
  core solution does not intersect all the contours for an envelope
  with polytropic index $3$. We find that this analysis also applies
  to similar limits in the literature including the inner mass limit
  for polytropic models of quasi-stars.  Consequently, any core
  solution that does not intersect all the fractional mass contours
  exhibits an associated limit and we identify several relevant cases
  where this is so.  We show that a composite polytrope is at a
  fractional core mass limit when its core solution touches but does
  not cross the contour of the corresponding fractional core mass. We
  apply this test to realistic models of helium stars and find that
  stars typically expand when their cores are near a mass
  limit. Furthermore, it appears that stars that evolve into giants
  have always first exceeded an SC-like limit.
\end{abstract}

\begin{keywords}
  stars: evolution -- stars: interiors
\end{keywords}

\section{Introduction}

Once the core of a main-sequence star has exhausted its supply of
hydrogen, it ceases to produce nuclear energy and, in the limit of
thermal equilibrium, becomes isothermal. \citet{sc42} showed that, if
the envelope is polytropic with index $n=3$, then there is a maximum
fractional mass that the core can achieve. If the core is less
massive, it can remain isothermal while nuclear reactions continue in
a surrounding shell. If this mass is exceeded then the core contracts
until it is supported by electron degeneracy pressure or helium begins
to burn at the centre. The idealised result is sufficiently accurate
that it has become a well-established element of the theory of the
post-main sequence evolution of stars. It is referred to simply as the
Sch\"onberg--Chandrasekhar (SC) limit.

Similar limits have been computed for other polytropic solutions.
\citet{beech88} calculated the corresponding limit for an isothermal
core surrounded by an envelope with \mbox{$n=1$}.  \citet*{efc98}
found that, when \mbox{$n=1$} in the envelope and \mbox{$n=5$} in the
core, a fractional mass limit exists if the density decreases
discontinuously at the core-envelope boundary by a factor exceeding
3. They went further to propose conditions on the polytropic indices
of the core and envelope that lead to fractional mass limits. We refer
to all these limits, including the original result of \citet{sc42} as
\emph{SC-like} limits.

Previously, we found that the black hole mass of a polytropic
quasi-star exhibits a robust fractional limit \citep{ball+11}.  We
have determined why this limit exists in terms of contours of
fractional core mass of solutions when plotted in the space of
homology invariant variables $U$ and $V$.  We have further found that
all SC-like limits are explained by a similar approach. In this work,
we present our analysis, which unifies SC-like limits and indicates
that they exist in a wider range of circumstances than the handful of
cases discussed in the literature. In Section \ref{suvp}, we provide a
thorough exposition of the relevant features of the \uvp{}. Readers
who are familiar with these details can proceed to Section
\ref{scont}, where we present our new interpretation of SC-like
limits. In Section \ref{sgen}, we provide a description that captures
all the SC-like limits of Section \ref{scont}. We also consider how to
determine whether a star has reached an SC-like limit and how the
fractional mass contours constrain its evolution and we conclude in
Section \ref{sconc}.

\section{The $U$--$V$ plane}
\label{suvp}

This work rests on the behaviour of solutions in the \uvp{} so we
begin with a review of its features. We derive the Lane--Emden
equation (LEE) from hydrostatic equilibrium and mass conservation,
introduce homology invariant variables $U$ and $V$ and explain their
physical meaning, present the homology invariant transformation of the
equation and study the behaviour of its solutions in the \uvp{}.  We
hope that, by presenting concisely the details of the \uvp{} in a
context where it is usefully applied, we might remove its stigma as
`that gruesome tool'.\footnote{\citet{faulkner05} explains that Martin
  Schwarzschild described the \uvp{} as such in a referee's report in
  1965. The same quote is presumably the citation by \citet{efc98} of
  `(Schwarzschild 1965, private communication)'.}

\subsection{The Lane-Emden Equation}

Consider the equations of mass conservation,

\shorteq{\label{dmdr}\frac{dm}{dr}=4\pi r^2\rho\text{,}} and
hydrostatic equilibrium,

\shorteq{\label{dpdr}\frac{dp}{dr}=-\frac{Gm\rho}{r^2}\text{,}} where
$r$ is the distance from the centre of the star, $m$ is the mass
within a concentric sphere of radius $r$, and $p$ and $\rho$ are the
pressure and density, respectively. We make the usual polytropic
assumption that the pressure and density are related by
$p=K\rho^{1+\frac{1}{n}}$, where $n$ is the polytropic index and $K$
is a constant of proportionality.  We define the dimensionless
temperature\footnote{This is by the analogy to an ideal gas, for which
  $T\propto p/\rho$.}  $\theta$ by $\rho=\rho_c\theta^n$, where
$\rho_c$ is the density at the centre of the star, the dimensionless
radius\footnote{The scale factor is usually denoted $\alpha$. We have
  used $\eta$ to avoid confusion with the density jump at the
  core-envelope boundary, which \citet{efc98} denoted $\alpha$.}
$\xi$ by $r=\eta\xi$, where

\shorteq{\eta^2=\frac{(n+1)K}{4\pi G}\rho_c^{\frac{1}{n}-1}\text{,}}
and the dimensionless mass $\phi$ by $m=4\pi\eta^3\rho_c\phi$.

We use a polytropic equation of state to approximate a fluid that is
between the adiabatic and isothermal limits. Shallower temperature
gradients correspond to larger effective polytropic indices and the
isothermal case (zero temperature gradient) corresponds to
$n=\infty$. In this case, the equation of state must be approximated
differently but the limit is well-defined when working in the \uvp{}.

Certain conditions inside a star correspond to certain values of
$n$. In convective zones, the temperature gradient is approximately
adiabatic, so an ideal gas without radiation has $n=3/2$ and pure
radiation has $n=3$. Real stars are more radiation-dominated towards
the centre and $n$ varies between these limiting values in convection
zones. In radiative zones, $n$ depends on the opacity.  It can be
shown, for example, that for a polytropic model with uniform energy
generation and a Kramer's opacity law, $n$ ranges from $13/4$ for a
pure ideal gas to $7$ for pure radiation \citep{horedt04}.  Nuclear
burning shells and ionisation regions have shallow temperature
gradients and therefore large values of $n$. Thus, the effective
polytropic index can vary widely within a star.

Introducing the dimensionless mass, temperature and radius into
equations (\ref{dmdr}) and (\ref{dpdr}) allows us to write

\shorteq{\label{dphdxi}\frac{d\phi}{d\xi}=\xi^2\theta^n} and

\shorteq{\label{dthdxi}\frac{d\theta}{d\xi}=-\frac{1}{\xi^2}\phi\text{.}}
By differentiating equation (\ref{dthdxi}) and substituting for
$d\phi/d\xi$ from equation (\ref{dphdxi}), 
we arrive at the LEE,

\shorteq{\label{LE}\frac{1}{\xi^2}\frac{d}{d\xi}
  \left(\xi^2\frac{d\theta}{d\xi}\right)=-\theta^n\text{,}}  
in its usual form as a single second-order ordinary differential
equation. Here, we prefer to express it as two first-order equations
(\ref{dphdxi} and \ref{dthdxi}) because this preserves the physical
meaning of the equations and easily permits arbitrary boundary
conditions for the inner mass and radius. 

Solutions of the LEE which are regular at the centre have
$\xi_c=\phi_c=0$. The subset of solutions that extend from the centre
to infinite radius or the first zero of $\theta$ are \emph{polytropes}
of index $n$. We refer to solutions that are regular at the centre but
truncated at some finite radius as \emph{polytropic cores}.
Conversely, solutions that extend from a finite radius to infinity or
the first zero of $\theta$ are \emph{polytropic envelopes}. In
addition, we refer to models that match polytropic cores to polytropic
envelopes as \emph{composite} polytropes.  For $n<5$ polytropes are
finite in both mass and radius while for $n>5$ they are infinite in
mass and radius. The case $n=5$ represents the threshold between the
two: it has a finite mass but infinite radius.

\subsection{Homology Invariant Variables}

It is known \citep[see][]{chandra39} that, if $\theta(\xi)$ is a
solution of the LEE, then $\theta'(\xi')=C^\frac{2}{n-1}\theta(C\xi)$,
where $C$ is an arbitrary constant, is also a solution. The two
solutions are \emph{homologous}; the similarity between them is called
\emph{homology}. By choosing variables that are invariant under this
transformation, we can formulate the LEE as a single first-order
equation that captures all essential behaviour.
We use the variables

\shorteq{\label{Ugen}U=\frac{d\log m}{d\log r}=\frac{3\rho}{\bar\rho}}
and

\shorteq{\label{Vgen}V=-\frac{d\log p}{d\log r}
  =\frac{Gm}{r}\frac{\rho}{p}\text{,}} 
where $\bar\rho=3m/4\pi r^3$ is the mean density of the material
inside $r$. Although we have defined $U$ and $V$ to reduce the order
of the LEE, the corresponding physical definitions make them
meaningful for discussions of any stellar model.  We make a brief
excursion to explain these definitions.

\begin{figure}
  \includegraphics[width=84mm]{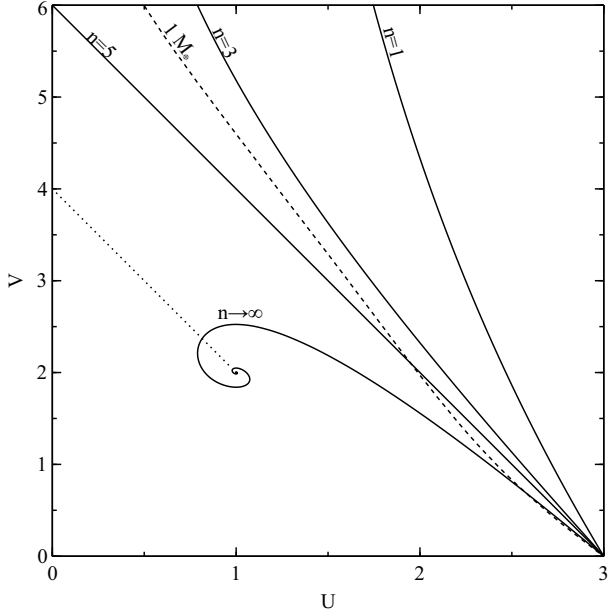}
  \caption{Some general features of the \uvp{}. The solid lines are,
    from top to bottom, polytropes of index $1$, $3$, $5$ and
    $\infty$. The dashed line is a \stars{} model of a $1\Msun$ star
    of solar metallicity when its radius is $1.012\Rsun$ and its
    luminosity $0.974\Lsun$.  The dotted line shows the locus of the
    critical points $G_s$. The locus begins in the plane at $(0,4)$ when
    $n=3$ and tends to $(1,2)$ as $n$ increases to $\infty$.}
  \label{fgen}
\end{figure}

The physical variables are all positive so only the first quadrant
($U,V>0$) of the \uvp{} is of interest. The variable $U$ is three
times the ratio of the local density to the mean density inside that
radius. As $r\to0$, we also have $\rho\to\bar\rho$ and thus $U\to3$.
We expect that the density of a stellar model decreases with radius,
so $\rho/\bar\rho<1$ and hence $U<3$. The variable $V$ is related to
the ratio of specific gravitational binding energy to specific
internal energy. At the centre, $p$ and $\rho$ are finite and $m\sim
\frac{4\pi}{3}\rho r^3$, so $V\to0$. Thus, in all physical solutions
that extend to $r=0$, the centre corresponds to $(U,V)=(3,0)$. If an
interior solution has $V$ everywhere smaller than the appropriate
polytrope then it behaves as if it has a finite point mass at its
centre. \citet{hs75} referred to similar models, integrated outwards
from a finite radius, as \emph{loaded polytropes}. If a solution has
$V$ everywhere greater than the polytrope then it reaches zero mass
before zero radius.  Such solutions would have a massless core with
finite radius, which is unphysical. At the surface, $\rho\to0$ so
$U\to0$ too. On the other hand, $Gm/r$ takes a finite value but
$p/\rho\propto T\to0$ so $V\to\infty$. All realistic models, be they
polytropes, composites of a polytropic core and envelope, or output
from a detailed calculation, must adhere to these central and surface
conditions in the \uvp{}. They therefore extend from $(3,0)$ towards
$(0,\infty)$.  Fig.\,\ref{fgen} shows this behaviour for polytropes of
indices $1$, $3$, $5$ and $\infty$. The $n=1$ and $n=3$ models extend
properly to the surface. The $n=5$ and $\infty$ polytropes do not and
therefore cannot represent a real star. In addition, we have plotted a
$1\Msun$ model produced by the Cambridge \stars{} code to show that it
also satisfies the boundary conditions described above. Note that we
have not calibrated this model to fit the Sun precisely.

For a composite polytrope, the pressure, mass and radius are
continuous at the join. If the density is decreased by a factor
$\alpha$ \citep[c.f.][]{efc98}, then $U$ and $V$ decrease by the same
factor. In other words, if $\rho\to\alpha^{-1}\rho$ then
$(U,V)\to\alpha^{-1}(U,V)$. The corresponding point on the \uvp{} is
contracted towards the origin by the factor $\alpha$. Such a jump
occurs, for example, if there is a discontinuity in the mean molecular
weight $\mu$ between the core and the envelope.  In this case,
$\alpha=\mu_c/\mu_e$.

Let us now return to the polytropic solutions for which we defined $U$
and $V$ in the first place. From the definitions above,

\shorteq{\label{Upol}U=\frac{d\log\phi}{d\log\xi}=\frac{\xi^3\theta^n}{\phi}}
and

\shorteq{\label{Vpol}V=-(n+1)\frac{d\log\theta}{d\log\xi}
  =(n+1)\frac{\phi}{\theta\xi}\text{.}}  Let us differentiate $\log U$
and $\log V$ as they are defined for polytropes. This gives

\shorteq{\label{dU}\frac{1}{U}\tdif{U}{\xi}=\frac{1}{\xi}[3-n(n+1)^{-1}V-U]}
and
\shorteq{\label{dV}\frac{1}{V}\tdif{V}{\xi}=\frac{1}{\xi}[-1+U+(n+1)^{-1}V]\text{.}}
The ratio of these two equations yields the first-order equation

\shorteq{\label{HLEE}\tdif{V}{U}=-\frac{V}{U}
  \left(\frac{U+(n+1)^{-1}V-1}{U+n(n+1)^{-1}V-3}\right)} in which the
dependence on $\xi$ has been eliminated. We refer to equation
(\ref{HLEE}) as the homologous Lane--Emden equation (HLEE). The
SC-like limits we wish to reproduce are shared by polytropic models so
we now explore the behaviour of these solutions in the plane defined
by $U$ and $V$.

\subsection{Topology of the Homologous Lane--Emden Equation}
\label{sstop}

The behaviour of solutions of the HLEE is described in terms of its
critical points, where $dU/d\log\xi$ and $dV/d\log\xi$ both tend to
zero.  \citet{horedt87} conducted a thorough survey of the behaviour
of the HLEE, including the full range of $n$ from $-\infty$ to
$\infty$ in linear, cylindrical and spherical
geometries.\footnote{Readers should note that the definition of $V$
  used by \citet{horedt87} is smaller by a factor $n+1$.} Below, we
use his convention for naming the critical points but consider only
spherical cases with $n\geq1$.  Though realistic polytropes take $n$
in the range $3/2$ to infinity, we extend it to accommodate SC-like
limits discussed in the literature for polytropic envelopes with
$n=1$.

From the numerator of equation (\ref{HLEE}), we see that $dV/dU=0$
when $V=0$ or $U+V/(n+1)=1$. The former indicates that solutions that
approach the $U$-axis proceed along it until they reach infinity or a
critical point. The latter defines a straight line in the \uvp{} along
which solutions are locally horizontal. Following \citet{faulkner05}
we refer to this line as the \emph{line of horizontals}. Similarly,
from the denominator, we find $dU/dV=0$ when $U=0$ or $U+nV/(n+1)=3$.
Again, the first locus implies that solutions near the $V$-axis have
trajectories that are nearly parallel to it, while the second gives
another straight line, this time along which solutions are vertical,
hereinafter referred to as the \emph{line of verticals}.  The critical
points of the HLEE are located at the intersections of these curves.
Below, we consider the stability of the critical points as $n$
varies. The analysis on which the discussion is based is provided in
the Appendix.

The origin is the first critical point. It is a saddle with the
solutions on the $V$-axis approaching and those on the $U$-axis
escaping. Solutions near the origin move down and to the right on the
\uvp{}.
There is a further critical point on each of the axes. On the
$U$-axis, $U_s=(3,0)$ is a saddle for all values of $n$. It is stable
along the $U$-axis and unstable across it. This point coincides with
the regular centre of realistic stellar models that we discussed
previously. Along the $V$-axis, $V_s=(0,n+1)$ is also a critical
point. For $n<3$ it is a source and for $n>3$ a saddle.
The intersection of the lines of horizontals and verticals is the 
final critical point $G_s$. For each n,
$G_s=(\frac{n-3}{n-1},2\frac{n+1}{n-1})$. 
The character of these points varies with $n$. Their locus is shown in
Fig.\,\ref{fgen}.

\begin{figure}
  \includegraphics[width=84mm]{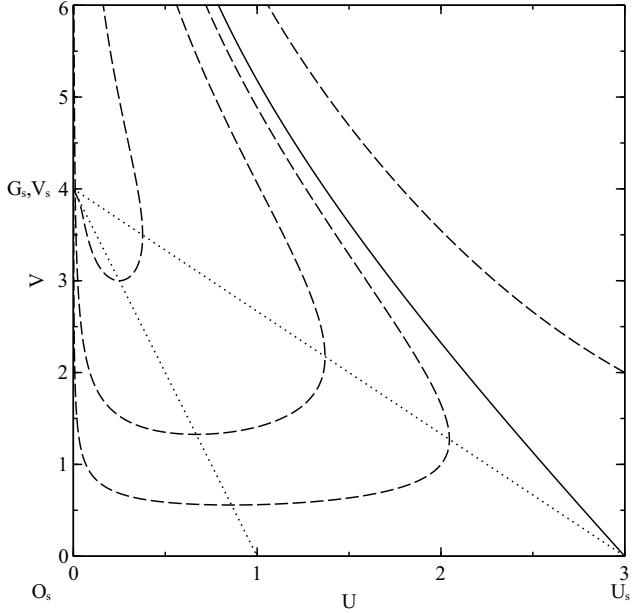}
  \caption{Topology of the HLEE for $n=3$. The solid line is the
    polytrope. The upper and lower dotted lines are the lines of
    verticals and horizontals, respectively. The dashed lines are a
    selection of solutions. Below the polytrope, the solutions have a
    non-zero mass at the centre, which is represented by the critical
    point $V_s=(0,4)$. Conversely, solutions above the polytrope have
    zero mass at non-zero inner radius.}
  \label{ftop3}
\end{figure}

The behaviour of $G_s$ and $V_s$ distinguishes the topology of
solutions into three regimes. For $n<3$, $V_s$ is a pure source: it is
unstable across and along the $V$-axis. The point $G_s$ has $U<0$ and
therefore does not feature in the first quadrant of the \uvp{} but
approaches the $V$-axis from the left as $n\to3$. When $n=3$, $V_s$
and $G_s$ co-incide. The point is marginally stable across the
axis. For $n>3$, $V_s$ and $G_s$ separate. $V_s$ is now a saddle and
$G_s$ a source, gradually moving towards its position at $(1/2,3)$
when $n=5$. Fig.\,\ref{ftop3} illustrates some features of the HLEE
when $n=3$. The lines of verticals and horizontals meet at $G_s$,
which has just appeared on the \uvp{} at $(0,4)$.

\begin{figure}
  \includegraphics[width=84mm]{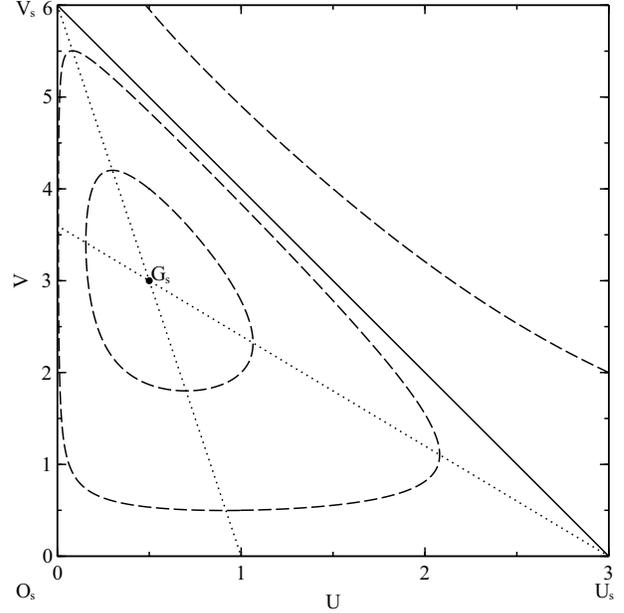}
  \caption{Topology of the HLEE for $n=5$. The solid line is the
    polytrope. The dotted line from $V_s=(6,0)$ to $(1,0)$ is the line
    of horizontals. The other dotted line, from $(0,18/5)$ to $(3,0)$,
    is the line of verticals. The point $G_s=(1/2,3)$ is now distinct
    from $V_s$ and is a centre, as can be seen from the solutions,
    represented by the dashed lines. Below the polytrope, the
    solutions orbit $G_s$. Above, they extend from a finite inner
    radius, where $\phi=0$, to a finite outer mass and radius.}
  \label{ftop5}
\end{figure}

When $n=5$, which separates the cases of finite and infinite
polytropes, the \uvp{} takes on a particular structure, illustrated in
Fig.\,\ref{ftop5}. The $n=5$ polytrope is a straight line from
$U_s=(3,0)$ to $V_s=(0,6)$. The point $G_s$ is a centre, with
solutions forming closed loops around it.  The polytrope separates
solutions that circulate around $G_s$ from those that go from
$(\infty,0)$ to $(0,\infty)$ entirely above the polytrope. These
solutions have zero mass at non-zero inner radius but, unlike the
polytrope, have a finite outer radius.

As $n$ increases further $G_s$ becomes a spiral sink. Polytropes start
at $U_s$ and now spiral into $G_s$ (see Fig.\,\ref{fgen}). There is an
unstable solution that proceeds from $(\infty,0)$ to $V_s$ above which
solutions extend to $(0,\infty)$. As $n\to\infty$ we also find
$V_s\to(0,\infty)$ and, in the limiting case of the isothermal sphere,
\emph{all} solutions ultimately spiral into $G_s$ because they cannot
lie above the unstable solution.

\section{Fractional core mass contours}
\label{scont}

Let us consider the problem of fitting a polytropic envelope to a core
of arbitrary mass and radius.  For a given $n<5$, we can regard a
given point $(U_0,V_0)$ in the \uvp{} as the interior boundary of a
corresponding polytropic envelope by integrating the LEE from that
point to the surface. More precisely, we can take interior conditions

\shorteq{\theta_0=1\text{,}}
\shorteq{\xi_0=\sqrt{(n+1)^{-1}U_0V_0}\text{,}} and
\shorteq{\phi_0=\sqrt{(n+1)^{-3}U_0V_0^3}}
and integrate the LEE up to the first zero of $\theta$ where we set
$\xi=\xi_1$. This point marks the surface of a polytropic envelope, at
which the dimensionless mass co-ordinate $\phi_1$ is the total mass of
the solution, including the initial value $\phi_0$. The ratio
$q=\phi_0/\phi_1$ is then the fractional mass of a core that occupies
a dimensionless radius $\xi_0$.  By associating each point in the
\uvp{} with the value of $q$ for a polytropic envelope that starts
there, we define a surface $q(U,V)$.  We use the contours of this
surface to characterise the SC limit.

\begin{figure}
  \includegraphics[width=84mm]{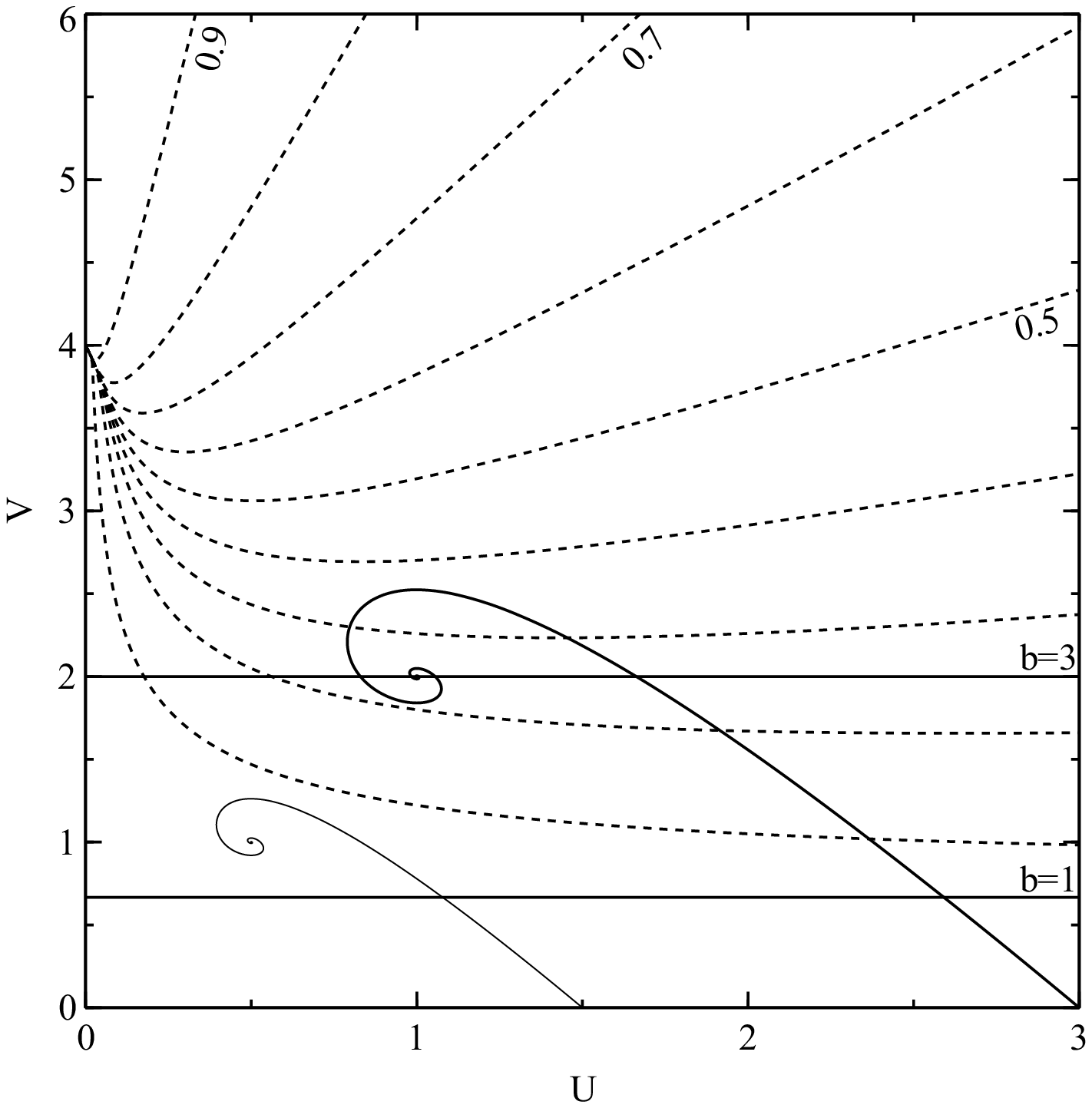}
  \caption{The dashed lines are contours of a core's fractional mass
    $q=\phi_0/\phi_1$ beneath an envelope with $n=3$. They increase in
    steps of $0.1$ from $0.1$, at the bottom, to $0.9$, at the top.
    The larger solid spiral is the isothermal core with
    $\alpha=1$. The smaller spiral represents an isothermal core when
    $\alpha=2$. The upper and lower straight lines represent the inner
    boundaries for quasi-stars with $b=3$ and $b=1$ respectively 
    (see Section \ref{ssqs}).}
  \label{fcont3}
\end{figure}

\begin{figure}
  \includegraphics[width=84mm]{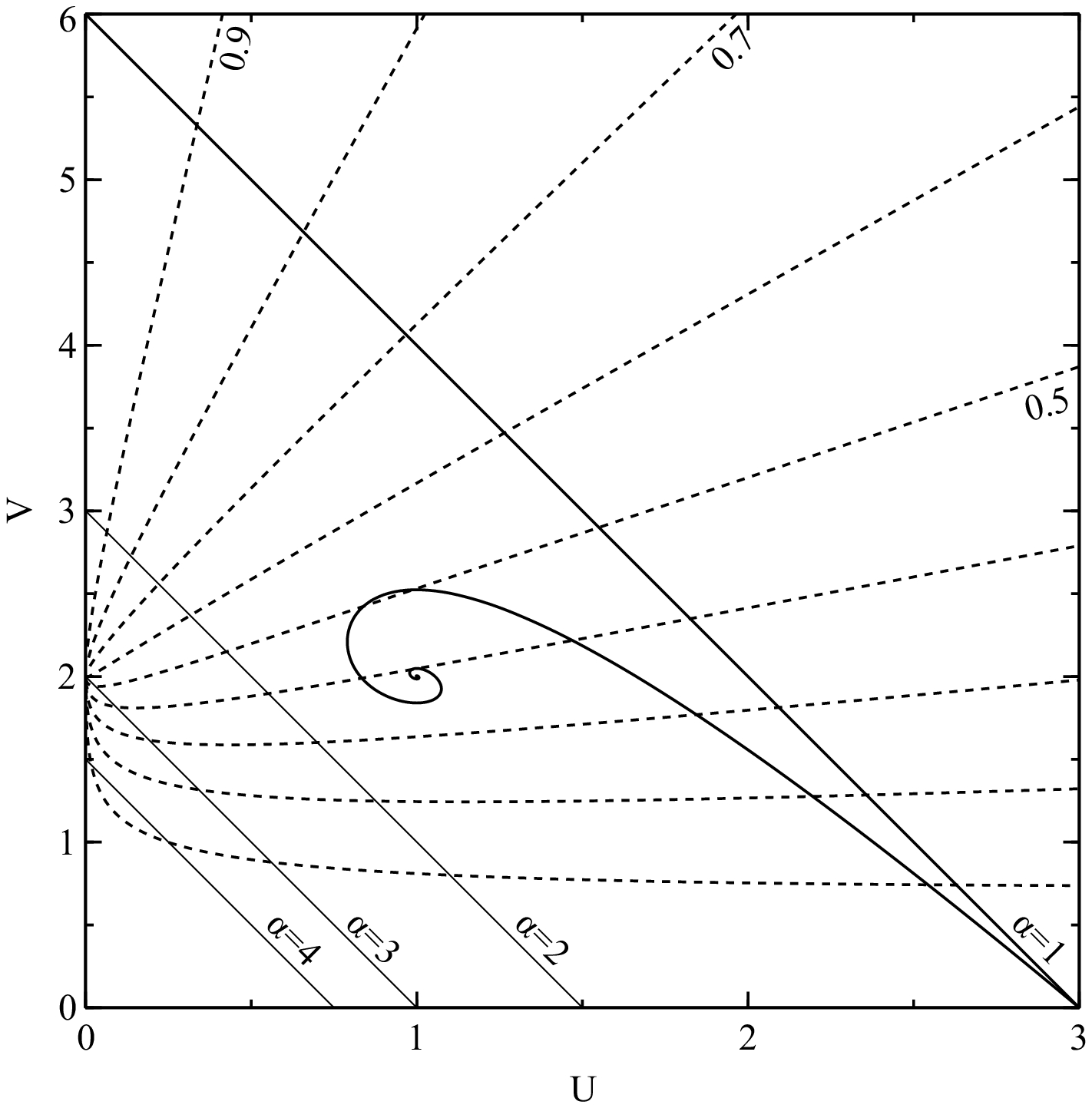}
  \caption{The dashed lines are contours of fractional mass
    $q=\phi_0/\phi_1$ for $n=1$. The solid spiral is again the
    isothermal core solution. The top-most diagonal line is the
    polytrope of index $5$ with $\alpha=1$. The other diagonal lines
    are, from top to bottom, core-envelope boundary conditions for the
    envelope when $\alpha=2,3,4$ for $n=5$ as shown by
    \citet[][Fig.\,3]{efc98}.}
  \label{fcont1}
\end{figure}

Figs\,\ref{fcont3} and \ref{fcont1} show contours of $q(U,V)$ for
polytropic envelopes with $n=3$ and $n=1$ respectively, along with a
selection of interior solutions that lead to SC-like limits. For $n<3$
the contours are dominated by the critical point $V_s$ and for $n>3$
by $G_s$. Away from $V_s$ or $G_s$ all the contours at first curve
away from the $U$-axis and then tend towards straight lines.

\subsection{The Sch\"onberg--Chandrasekhar Limit}

\citet{kw90} discuss the SC limit in terms of fractional mass
contours. \citet{cannon92} also explicitly described the SC limit in
terms of fractional mass contours, although he employed a different
set of homology-invariant variables.  Fig.\,\ref{fcont3} shows the
isothermal solution along with the fractional mass contours for $n=3$
envelopes. The SC limit exists because the isothermal solution only
intersects fractional mass contours up to a maximum
$q_\text{max}=0.359$ when $\alpha=1$. In other words, along the
isothermal solution, the function $q(U,V)$ achieves a maximum of
$0.359$.

A maximum $V$ for the isothermal curve exists because, for a given
mass, there is a finite maximum pressure that a core can exert.  This
limit is usually derived by defining the core pressure using virial
arguments and maximizing it with respect to the core radius
\citep[e.g.][]{kw90}. Such an explanation partly describes the SC
limit but our interpretation makes clear that the existence of the SC
limit has as much to do with the behaviour of the envelope solutions
as the isothermal core. For example, changing the polytropic index of
the envelope changes the mass limit.

If there is a density jump by a factor $\alpha$ at the core-envelope
boundary (see Section 2.2), then $U$ and $V$ at the edge of the core
must be transformed to find the base of the envelope in the \uvp{}.
That is, if $\rho\to\alpha^{-1}\rho$, then $(U,V)\to\alpha^{-1}(U,V)$.
The contraction of the isothermal core for $\alpha=2$ is included in
Fig.\,\ref{fcont3}. The inner boundary of the envelope shifts to a
smaller fractional mass of about $0.09$ so the SC limit falls too.

The argument presented here implies that SC-like limits exist whenever
an envelope is matched to a core that only intersects fractional mass
contours of that envelope up to some maximum. We now use this to
explain the existence of other mass limits in the literature.

\subsection{Related Polytropic Limits}

Fig.\,\ref{fcont1} shows the fractional mass contours for $n=1$, the
isothermal solution, and $n=5$ polytropes with $\alpha=1,2,3,4$ as
used by \citet{efc98}.  \citet{beech88} calculated an SC-like limit
for an isothermal core embedded in a polytropic envelope with $n=1$.
Because the behaviour of fractional mass contours is similar for $n=1$
and $n=3$, the existence of the limit is now no surprise. Note that
the numerical value found by \citet{beech88} differs because he
included the radiation pressure of the isothermal core, which
displaces $U$ and $V$ at the core-envelope boundary.

The conclusions of \citet{efc98} are also catered for. The critical
point $V_s=(0,n+1)$ separates solutions, and thus contours, with
$q\approx0$ from those with $q\approx1$. \citet{efc98} concluded that,
for $n=1$ envelopes, cores with $n<5$ are never subject to an SC
limit; those with $n>5$ always are; and those with $n=5$ constitute
the marginal case for which the limit exists when $\alpha>3$.
\citet{efc98} extended their arguments to changing the polytropic
index of the envelope. Because these conclusions are based on the
critical behaviour of the solutions, which is reflected in the
behaviour of the contours, the same results follow here. We have shown
how they are characterised by the contours in the same way as other
limits and are a particular example of our broader result. That is, we
have shown that SC-like limits exist whenever the core solution fails
to intersect all fractional mass contours. The cases identified by
\citet{efc98} fall within this description.

\subsection{Loaded Polytropes and Quasi-stars}
\label{ssqs}

Quasi-stars are objects consisting of a stellar-mass black hole
embedded in a massive, hydrostatic, giant-like envelope, potentially
formed when primordial gas in large dark matter haloes collapsed in
the early Universe \citep{bvr06}.  The black hole is able to grow
rapidly as long as the hydrostatic structure persists. Following the
simple models described by \citet*{bra08}, \citet{ball+11} computed
models with the Cambridge \stars{} code and found a maximum mass for
the black hole that was accurately reproduced by polytropic models. We
now show how this result is related to our analysis of the SC limit.

The interior boundary condition for the quasi-star models can be
written as

\shorteq{r_0=\frac{1}{b}\frac{2Gm_0}{c_s^2}\label{r0def}\text{,}}
where $b$ is a scale factor, $m_0$ is the mass interior to $r_0$ and
$c_s^2=\gamma p/\rho$ is the adiabatic sound speed. The boundary
condition is then a fraction $1/b$ of the Bondi radius, where
$mc_s^2/2=Gm/r$. \citet{bra08} used $b=3$; \citet{ball+11} used
$b=1$. Accretion on to the central black hole supports the envelope by
radiating near the Eddington limit of the entire object so the
envelope is strongly convective and the pressure is dominated by
radiation. The envelope is approximately polytropic with index $n=3$.
Now, at the interior boundary, $c_s^2=(1+1/n)K\rho_c^{1/n}$,
$m_0=4\pi\eta^3\rho_c\phi_0$, and $r_0=\eta\xi_0$ so

\shorteq{\label{qbc}\phi_0=\frac{b}{2n}\xi_0\text{.}} 
Transforming to $U$ and $V$ gives $U_0=2n\xi_0^2/b$ and
$V_0=b(n+1)/2n$. Varying $\xi_0$ traces a straight line, parallel to
the $U$-axis. In Fig.\,\ref{fcont3}, we have plotted $V_0=2/3$ and $2$,
which correspond to $b=1$ and $3$, respectively, for $n=3$. The line
of $V_0$ does not intersect all the contours of fractional core mass
because many of them are positively curved. Thus, a mass limit exists,
as in previous cases. For larger values of $b$, $V_0$ is also larger
and intersects more of the contours. The mass limit is therefore
larger.

We have limited ourselves to the case where $n=3$. The fractional mass
contours in Figs\,\ref{fcont3} and \ref{fcont1} show similar
behaviour.  Convective envelopes are approximately adiabatic and have
effective polytropic indices between $3/2$ and $3$, depending on the
relative importance of gas and radiation pressures. All such envelopes
possess fractional mass contours that are similar to the two cases
here and we conclude that a fractional mass limit for the black hole
exists in all realistic cases. Envelopes with $3<n<5$ have more
complicated fractional mass contours so we cannot immediately draw
similar conclusions.

This mass limit is not exactly the same as found by \citet{ball+11}.
They computed the Bondi radius using the mass of the black hole only,
even once the mass of gas inside the Bondi radius is comparable to
(and even exceeds) the mass of the black hole. Begelman (2010, private
communication) pointed out that the Bondi radius should be defined for
the total mass inside $r_0$, not just the black hole mass. Presuming
that the gas has a density distribution $\rho(r)\propto r^{-3/2}$
inside the cavity around the black hole, this gives the equation

\shorteq{r_0=\frac{2G}{c_s^2}
  \left(\Mbh+\frac{8\pi}{3}\rho_0r_0^3\right)\text{,}} where $\Mbh$ is
the mass of the black hole only, for $b=1$. Making the same
substitutions as in equation (\ref{qbc}) for polytropic index $n=3$,
the equation becomes

\shorteq{\phi_\text{BH}+\frac{2}{3}\xi_0^3-\frac{1}{6}\xi_0=0\text{,}}
which only has a real positive root if $\phi_\text{BH}<1/(18\sqrt{3})$.
The corresponding fractional core mass limit is $q=0.0166$.

In trying to move $r_0$ inwards, we found \citep{ball+11} we could not
construct models with $b\geq3.8$ in equation (\ref{r0def}). We can see
how this comes about from the behaviour of the polytropic limit. As
$b$ increases, $V_0$ increases and eventually passes the critical
point $G_s$ when the nature of the limit is reversed. When $b>2n$,
small inner masses correspond to envelopes with negligible envelope
mass so it becomes impossible to embed a small black hole inside a
massive envelope. In other words, the mass limit becomes a minimum
inner mass limit. For the models of \citet{ball+11}, the finite mass
of the black hole corresponds to a finite value of $U_0$ that
displaces the envelope slightly from the $V$-axis. The fractional mass
contours are closely packed near $G_s$, so a small value of $U_0$
introduces a minimum inner mass limit for $b<2n$. 

\section{General Limits}
\label{sgen}

The SC-like limits discussed above all exist because each locus of
core-envelope boundaries only intersects fractional mass contours with
$q$ smaller than some $q_\text{max}$. We can use this condition to
identify large classes of core solutions that lead to SC-like
limits. For example, an SC-like limit must exist whenever the core
solution has everywhere $V<n_\text{e}+1$, where $n_\text{e}<5$ is the
polytropic index of the envelope. Cores described by $n=5$ polytropes
with $\alpha>6/(n_\text{e}+1)$, as discussed by \citet{efc98}, satisfy
this condition. SC-like limits also exist whenever the curve defining
the inner edge of the envelope touches but does not cross some
fractional mass contour.  This explains the original SC limit and also
implies that any composite polytrope with $n>5$ in the core and $n<5$
in the envelope is subject to an SC-like limit. In fact, the limits
are determined by behaviour of solutions at the core-envelope
boundary, so SC-like limits also exist when there is a layer with
$n\gg5$ at the base of the envelope.  The constraint becomes even
stronger as the density gradient at the core-envelope boundary becomes
steeper or the mean molecular weight jump becomes more pronounced.
Both of these conditions become relevant immediately after a star
leaves its core-burning sequence, which implies that SC-like limits
apply earlier in a star's life than previously thought.

For a given core solution and polytropic index of the envelope, there
are potentially two solutions either side of an SC-like limit that
correspond to the same fractional core mass. The evolutionary sequence
of static models determines which solutions occur in reality. For
example, quasi-stars are initially constructed by loading a star with
a small mass $q\ll10^{-3}$. These lie close to the $V$-axis so
solutions must start with $U_0\approx0$ on the locus of inner boundary
conditions and move towards greater $U_0$ as the black hole grows and
the quasi-star evolves. The evolution halts when the maximum mass is
achieved because the black hole cannot lose mass. Further solutions,
with smaller inner masses than the maximum, exist as $U_0$ increases
further. We have calculated polytropic models with $U_0$ greater than
the SC-limited value but we were not able to compute STARS models
along the same sequence.

It is also possible to test whether a given composite polytrope is at
an SC-like limit. If the core solution touches but does not cross the
contour corresponding to the given fractional core mass, then the
model is at an SC-like limit. If this condition is satisfied, then
extending or contracting the core can only admit a smaller fractional
core mass. The test can be applied to realistic stellar models but
identifying stars that have reached an SC-like limit is difficult.
The inner edge of the envelope is not clearly defined and the
effective polytropic index varies throughout the envelope.

\begin{figure}
  \includegraphics[width=84mm]{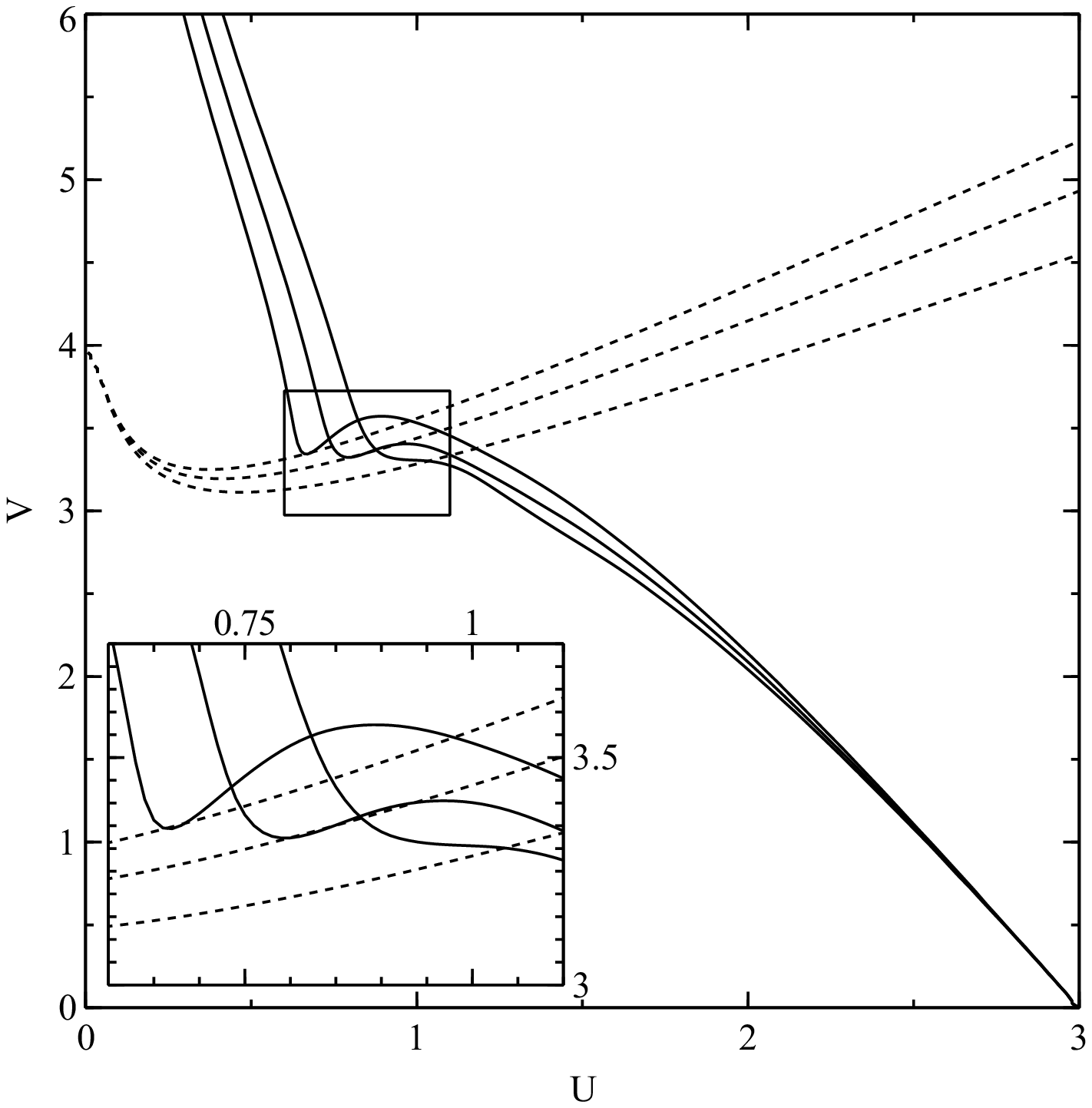}
  \caption{The solid lines are Cambridge \stars{} models of a $1\Msun$
    pure helium star.  Evolution proceeds from the right-most line at
    large $V$ to the left. The short-dashed lines are, from bottom to
    top, contours of fractional masses $0.516$, $0.543$ and $0.562$
    for $n=3$.  The inset is a magnification of the boxed region. The
    first model is at the end of the core He-burning sequence. The
    second model appears to have reached an SC-like limit of
    $q\approx0.543$ according to our criterion. At this point, the
    star begins to expand rapidly and $T_\text{eff}$ decreases.  As it
    evolves across the Hertzsprung gap, the gradient $\partial
    V/\partial U$ becomes steeper than the contours for the relevant
    fractional core mass.}
\label{fhe10}
\end{figure}

\begin{figure}
  \includegraphics[width=84mm]{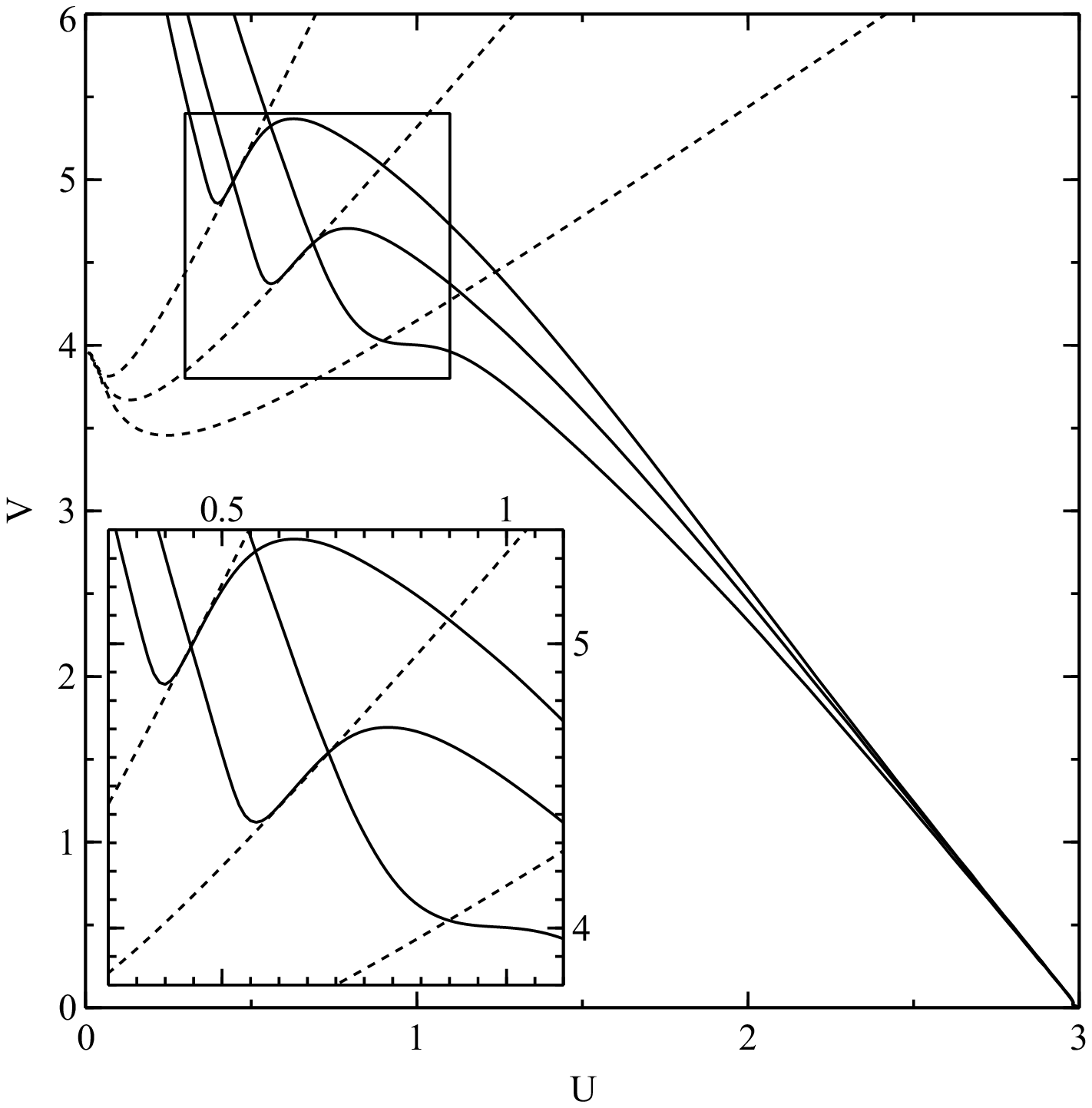}
  \caption{The solid lines are Cambridge \stars{} models of a
    $0.5\Msun$ pure helium star. Evolution proceeds from the
    right-most line at large $V$ to the left.  The short-dashed lines
    are, from bottom to top, contours of fractional masses $0.640$,
    $0.740$ and $0.825$ for $n=3$. The inset is a magnification of the
    boxed region.  The first model has already left the core
    He-burning sequence. The second model appears to be at an SC-like
    limit of $q_\text{max}\approx0.74$ but, although the star is
    expanding, it does not evolve into a giant. The third model is at
    the maximum luminosity achieved by the model and is still at or
    near an SC-like limit. As the star continues to evolve, its
    profile in the \uvp{} tends to a polytrope with $n\approx1.5$. The
    star does not become a giant and evolves directly on to the
    white-dwarf cooling sequence.}
  \label{fhe05}
\end{figure}

Despite the approximate nature of the test, we have applied it to
models produced by the Cambridge \stars{} code.  Figs\,\ref{fhe10} and
\ref{fhe05} show models of pure helium stars of $1\Msun$ and
$0.5\Msun$. The envelopes are still mostly radiative and their
effective polytropic indices vary between $n=2.5$ and $3.5$, so we
have used fractional mass contours for envelopes with $n=3$
everywhere. The mass co-ordinate at the core-envelope boundary was
determined approximately by eye at the point where $V$ was at a local
minimum. Where a minimum did not exist a similar nearby point in the
models was taken around where $\partial V/\partial U$ was a
maximum. Based on these, both stars appear to reach SC-like limits
shortly after moving off their core-burning sequences although their
subsequent evolution is different (see Section \ref{ssrg}). The regions
which lie parallel to the contours have $n\gg5$.  Although the $n=3$
contours are an approximation, it appears that both stars reach
SC-like limits shortly after leaving their core-burning sequence.

\subsection{Beyond the limit}

What happens when an isothermal core exceeds an SC-like limit? In
short, its effective polytropic index must change. This can happen in
two ways. Under suitable conditions, the inner part of the core
becomes degenerate. Degenerate matter is described by a polytropic
equation of state with $n=3$ or $n=3/2$ in relativistic and
non-relativistic cases, respectively. The inner core can tend to such
an equation with an isothermal layer further from the centre. An
SC-like limit still exists but the isothermal layer is displaced
upwards in $V$, so a larger core mass is possible. Alternatively, the
core departs from thermal equilibrium and contracts. For an ideal gas,
$dp/d\rho=1+1/n=1+dT/d\rho$, so the temperature gradient decreases the
effective polytropic $n$. As long as $n\gg5$, an SC-like limit
persists but, as in the previous case, it corresponds to a larger
fractional core mass.

The structure of the envelope offers some respite from the constraints
imposed by the core. For radiative envelopes, where $n$ is not much
greater than $3$, the fractional mass contours are much like those
shown in Fig.\,\ref{fcont3}. If the envelope becomes convective, then
the effective polytropic index varies between $3/2$ and $3$. An
SC-like limit still exists but the behaviour of the fractional mass
contours is less extreme near $V_s$ for smaller values of $n$ (compare
Figs \ref{fcont3} and \ref{fcont1}). Away from $G_s$, contours run
along smaller values of $V$ for smaller $n$. Equivalently, $q(U,V)$ is
larger at a given point $(U,V)$ for small values of $n$. For example,
for $n=3$, $q(2,4)=0.528$, whereas for $n=3/2$, $q(2,4)=0.562$ and for
$n=1$, $q(2,4)=0.576$. Thus, a smaller polytropic index in the
envelope permits a larger fractional core mass.

\subsection{Evolution into giants}
\label{ssrg}

Although the evolution of main-sequence stars into giants is
reproduced by detailed calculations of stellar structure, the cause of
a star's substantial expansion after leaving the main sequence remains
unknown. \citet{eggleton+91} showed that if the effective polytropic
index of a star is everywhere less than some $n_\text{max}<5$, then it
is less centrally condensed (i.e. $\rho_c/\bar\rho$ is smaller) than
the polytrope of index $n_\text{max}$. \citet{eggleton00} further
conjectured that the evolution of dwarfs into giants during shell
burning therefore requires that a significant part deep in the
envelope has an effective polytropic index $n\gg5$. This condition is
similar to the condition under which a star is subject to an SC-like
limit and the phenomena may be related.

Fig.\,\ref{fhe10} shows the evolution of a $1\Msun$ helium star. After
core He-burning is complete, the star briefly contracts and nuclear
fusion continues in the portion of the core that was not convective
during its burning phase. The star expands slightly before appearing
to reach an SC-like limit. At this point, the expansion of the
envelope accelerates and the effective temperature decreases,
indicating that it has begun to evolve into a giant. The star
continues to move across the Hertzsprung gap and the calculation is
terminated when C-burning begins at the centre of the core.  We also
evaluated the evolution of solar metallicity stars with masses $1$,
$3$, $5$, $7$ and $9\Msun$ and found qualitatively similar behaviour.
All appear to reach an SC-like limit at the end of the main sequence,
before starting to cross the Hertzprung gap and evolving into giants.

Fig.\,\ref{fhe05} shows the evolution of a $0.5\Msun$ helium star. The
star appears to reach an SC-like limit in its shell-burning phase when
the fractional core mass is about $0.74$.  The star expands briefly
but the $U$--$V$ profile tends back towards a polytrope thereafter. If
the mean molecular weight gradient were steeper then an SC-like limit
would apply earlier because the gradients of the contours are
shallower at smaller $V$.  However, the density jump between helium
and metals is modest and allows a large core to develop before a limit
applies.  The star's surface temperature does not decrease until it
reaches the white dwarf cooling sequence. Thus, although this star
appears to reach an SC-like limit, it does not become a giant.

Both these stars appear to reach SC-like limits but only one becomes a
giant. What is the crucial difference between them? In the $0.5\Msun$
star, the expansion around the limiting point reduces the effective
polytropic index in the burning shells and the limit relaxes slightly.
In the $1\Msun$ star, the burning shell is thinner, hotter and deeper
within the envelope. The shell responds less to the expansion. It
appears that reaching the limit leads to expansion of the envelope but
the response of the structure may then release the star from its
SC-like limit as for the $0.5\Msun$ star and not the $1\Msun$.

Another difference between the two stars is that, following the
limiting point, the core of $1\Msun$ star extends beyond the SC-like
limit. The gradient of the core profile becomes steeper than the
fractional core-mass contour, unlike the $0.5\Msun$ star where it is
approximately equal and then becomes shallower. Such a structure is
not limited because, by reducing the radial size of the limiting
region in the \uvp{}, a larger fractional core mass can be
accommodated but in order to reach this point it must have been
limited before. 

\section{Conclusion}
\label{sconc}

We have shown that SC-like limits exist whenever the solution
describing a stellar core only intersects a fraction of contours of
constant fractional mass for envelopes with a given polytropic
index. Our description explains the original SC limit, SC-like limits
found by \citet{beech88} and \citet{efc98}, and the limit for
polytropic quasi-stars found by \citet{ball+11}. It also shows that
SC-like limits exist under a wide range of circumstances. This
includes models where the core solution touches but does not cross a
particular fractional mass contour, as is the case if there is a layer
with $n\gg5$ at the base of the envelope.

We also derived a test of whether a polytropic model is at an SC-like
limit.  If the core solution touches but does not cross the contour of
the appropriate fractional mass, then the model is
SC-limited. Although the condition is only approximate for realistic
models, we have applied it to helium stars and found that achieving a
limit corresponds with expansion of their envelopes. Stars that
clearly exceed an SC-like limit consistently evolve into giants but it
is not known if this connection is causal. We have thus demonstrated
that the original SC limit is a particular case of a broader
phenomenon, that SC-like limits apply earlier in a star's evolution
than previously thought, and that there is a connection between
exceeding these limits and evolving into a giant.

\section*{Acknowledgements}

WHB and AN\.Z are grateful to Ramesh Narayan for the discussion that
led to the authors pursuing this line of work. We also thank Peter
Eggleton for discussing the formation of giants. CAT thanks Churchill
College for a Fellowship.

{\small

\begin{thebibliography}{}

\bibitem[\protect\citeauthoryear{{Ball}, {Tout}, {{\.Z}ytkow} \&
  {Eldridge}}{{Ball} et~al.}{2011}]{ball+11}
{Ball} W.~H.,  {Tout} C.~A.,  {{\.Z}ytkow} A.~N.,    {Eldridge} J.~J.,  2011,
  \mnras, 414, 2751

\bibitem[\protect\citeauthoryear{{Beech}}{{Beech}}{1988}]{beech88}
{Beech} M.,  1988, \apss, 147, 219

\bibitem[\protect\citeauthoryear{{Begelman}, {Rossi} \& {Armitage}}{{Begelman}
  et~al.}{2008}]{bra08}
{Begelman} M.~C.,  {Rossi} E.~M.,    {Armitage} P.~J.,  2008, \mnras, 387, 1649

\bibitem[\protect\citeauthoryear{{Begelman}, {Volonteri} \& {Rees}}{{Begelman}
  et~al.}{2006}]{bvr06}
{Begelman} M.~C.,  {Volonteri} M.,    {Rees} M.~J.,  2006, \mnras, 370, 289

\bibitem[\protect\citeauthoryear{{Cannon}}{{Cannon}}{1992}]{cannon92}
{Cannon} R.~C.,  1992, PhD thesis, University of Cambridge

\bibitem[\protect\citeauthoryear{{Chandrasekhar}}{{Chandrasekhar}}{1939}]{chan%
dra39}
{Chandrasekhar} S.,  1939, {An Introduction to the Study of Stellar Structure}.
{Univ. Chicago Press, Chicago}

\bibitem[\protect\citeauthoryear{{Eggleton}}{{Eggleton}}{2000}]{eggleton00}
{Eggleton} P.~P.,  2000 Unsolved Problems in Stellar Evolution.
{Cambridge Univ. Press, Cambridge}, p.~172

\bibitem[\protect\citeauthoryear{{Eggleton} \& {Cannon}}{{Eggleton} \&
  {Cannon}}{1991}]{eggleton+91}
{Eggleton} P.~P.,  {Cannon} R.~C.,  1991, \apj, 383, 757

\bibitem[\protect\citeauthoryear{{Eggleton}, {Faulkner} \& {Cannon}}{{Eggleton}
  et~al.}{1998}]{efc98}
{Eggleton} P.~P.,  {Faulkner} J.,    {Cannon} R.~C.,  1998, \mnras, 298, 831

\bibitem[\protect\citeauthoryear{{Faulkner}}{{Faulkner}}{2005}]{faulkner05}
{Faulkner} J.,  2005 {The Scientific Legacy of Fred Hoyle}.
{Cambridge Univ. Press, Cambridge}, p.~149

\bibitem[\protect\citeauthoryear{{Horedt}}{{Horedt}}{1987}]{horedt87}
{Horedt} G.~P.,  1987, \aap, 177, 117

\bibitem[\protect\citeauthoryear{{Horedt}}{{Horedt}}{2004}]{horedt04}
{Horedt} G.~P.,  ed. 2004, Astrophysics and Space Science Library Vol.~306 of
  Astrophysics and Space Science Library.
{Kluwer, Dordrecht}

\bibitem[\protect\citeauthoryear{{Huntley} \& {Saslaw}}{{Huntley} \&
  {Saslaw}}{1975}]{hs75}
{Huntley} J.~M.,  {Saslaw} W.~C.,  1975, \apj, 199, 328

\bibitem[\protect\citeauthoryear{{Kippenhahn} \& {Weigert}}{{Kippenhahn} \&
  {Weigert}}{1990}]{kw90}
{Kippenhahn} R.,  {Weigert} A.,  1990, {Stellar Structure and Evolution}.
{Springer-Verlag, Berlin}

\bibitem[\protect\citeauthoryear{{Sch{\"o}nberg} \&
  {Chandrasekhar}}{{Sch{\"o}nberg} \& {Chandrasekhar}}{1942}]{sc42}
{Sch{\"o}nberg} M.,  {Chandrasekhar} S.,  1942, \apj, 96, 161

\bibitem[\protect\citeauthoryear{{Strogatz}}{{Strogatz}}{1994}]{strogatz94}
{Strogatz} S.~H.,  1994, {Nonlinear Dynamics and Chaos: with Applications to
  Physics, Biology, Chemistry and Engineering}.
{Perseus Books, Reading, MA}

\end{thebibliography}

}

\section*{Appendix A:}

From equations (\ref{dU}) and (\ref{dV}), we find

\shorteq{\tdif{U}{\log\xi}=-U[U+n(n+1)^{-1}V-3]} 
and
\shorteq{\tdif{V}{\log\xi}=V[U+(n+1)^{-1}V-1]\text{.}}
This is an \emph{autonomous} system of equations: the derivatives
depend only on the dependent variables $U$ and $V$. The linear
behaviour of such systems around the critical points can be
characterised by the eigenvectors and eigenvalues of the Jacobian
matrix \cite[e.g.][]{strogatz94},

\begin{align}J&=\left(\begin{array}{cc}
\pdif{}{U}\tdif{U}{\log\xi} & \pdif{}{V}\tdif{U}{\log\xi} \\
\pdif{}{U}\tdif{V}{\log\xi} & \pdif{}{V}\tdif{V}{\log\xi}
\end{array}\right)\\
&=\left(\begin{array}{c@{}c}
3-2U-\frac{n}{n+1}V & -\frac{n}{n+1}U \\ \\
V                   & -1+U+\frac{2}{n+1}V
\end{array}\right)\text{,}\end{align}
at the critical point in question. In particular, if the real
component of an eigenvalue is positive or negative, solutions tend
away from or towards that point along the corresponding
eigenvector. Such points are \emph{sources} or \emph{sinks}. When the
point has one positive and one negative eigenvalue it is a
\emph{saddle}. If the eigenvalues have imaginary components then
solutions orbit the point as they approach or recede. We describe
these as \emph{spiral} sources or sinks. If the eigenvalues are purely
imaginary, then solutions form closed loops around that point, which
we call a \emph{centre}. The choice of independent variable, in this
case $\log\xi$, is not relevant in such analysis.

Table 1 shows the eigenvalues and eigenvectors for the critical points
in Section \ref{sstop} as functions of $n$. The origin $O_s$ is always
a saddle, with paths approaching along the $V$-axis and escaping along
the $U$-axis.  Because $n\ge1$ for realistic or interesting models,
$U_s$ also keeps the same saddle behaviour in our discussion, with
points approaching along the axis and escaping along the other
eigenvector, which always points towards the top left of the
\uvp{}. For $n<3$, $V_s$ is a source. One eigenvector is always along
the $V$-axis and the other across it but the latter varies from
pointing up in the \uvp{} to pointing down. When $n=3$ one eigenvalue
is zero so that it is a point of marginal stability. Points along the
corresponding eigenvector are also stationary in the linear regime.
As $n$ increases past $3$, $V_s$ becomes a saddle, with points now
approaching from positive $U$.

Lastly, $G_s$ displays the most complicated behaviour. It first
appears in the \uvp{} when $n=3$. In this case, it co-incides with
$V_s$. As $n$ increases, $G_s$ is at first a source. When
$\Delta_n=\sqrt{1+n(22-7n)}=0$, the eigenvalues take on an imaginary
component, so $G_s$ becomes a spiral source. Increasing in $n$, the
special case $n=5$ is reached. The eigenvalues become purely imaginary
at $G_s$, so the point is a pure centre (see Fig.\,\ref{ftop5}). For
$n>5$, the real part of $G_s$ is negative and it becomes a spiral
sink. As $n\to\infty$, $V_s$ effectively vanishes and all solutions
ultimately reach $G_s$.

\begin{table}
\caption{Critical points of the HLEE. $\Delta_n=\sqrt{1+n(22-7n)}$.}
\begin{tabular}{cccccc}
\toprule
\mchead{Critical point} & \mchead{Eigenvalues} & \mchead{Eigenvectors}  \\
\midrule
$O_s$ & $(0,0)$         &   $3$     &  $-1$   & $(1,0)$ &   $(0,1)$    \\
$U_s$ & $(3,0)$         &   $-3$    &  $2$    & $(1,0)$ & $(-3n,5+5n)$ \\
$V_s$ & $(0,n+1)$       &   $1$     &  $3-n$  & $(0,1)$ & $(2-n,1+n)$  \\
$G_s$ & $\left(\frac{n-3}{n-1},2\frac{n+1}{n-1}\right)$ & 
 \mchead{$\frac{n-5\pm\Delta_n}{2-2n}$} & \mchead{$(1-n\mp\Delta_n,4+4n)$}  \\
\bottomrule
\end{tabular}
\end{table}

\end{document}